\documentstyle[12pt]{article}
\def\lesssim{{\
\lower-1.2pt\vbox{\hbox{\rlap{$<$}\lower5pt\vbox{\hbox{$\sim$}}}}\ }}
\def\gtrsim{{\
\lower-1.2pt\vbox{\hbox{\rlap{$>$}\lower5pt\vbox{\hbox{$\sim$}}}}\ }}

\begin{document}

\begin{titlepage}

\begin{flushright}
    UT-780
\end{flushright}

\vspace{0.5cm}

\begin{center}
    {\Large\bf Dynamical Tuning of the Initial Condition for New
    Inflation in Supergravity}\\
    \vspace{1.5cm}
    {\large Izawa K.-I.$^{a}$, M.~Kawasaki$^{b}$ and
    T.~Yanagida$^{a}$}\\
    \vspace{1cm}
    {\it $^{a}$Department of Physics, University of Tokyo, Tokyo 133,
    Japan\\
    $^{b}$Institute for Cosmic Ray Research, University of
    Tokyo, Tokyo 188, Japan}\\
   \vspace{0.5cm}
    July, 1997
\end{center}

\vspace{0.5cm}

\begin{abstract}
    We point out that for a class of `new inflation' models in
    supergravity the required initial value of the inflaton field is
    dynamically set if there is another inflation (`pre-inflation')
    before the `new inflation'. We study the dynamics of both
    inflatons by taking a hybrid inflation model for the
    `pre-inflation' as an example.  We find out that our `new
    inflation' model provides reheating temperatures $T_R \simeq
    10{\rm MeV}-10^4{\rm GeV}$ low enough to avoid the gravitino
    problem even in gauge-mediated supersymmetry-breaking models. We
    also construct a model where the scale for the `new inflation' is
    generated by nonperturbative dynamics of a supersymmetric gauge
    theory.
\end{abstract}


\end{titlepage}
\newpage


\section{Introduction}

The hypothesis of an exponentially expanding (inflationary) stage in
the early universe is a very attractive idea, since it solves the
flatness and the horizon problems~\cite{Guth} in the standard
cosmology.  There have been proposed various models to realize the
inflationary epoch in the early universe~\cite{Linde-book}.  Among
them the `new inflation' model~\cite{Albrecht} is the most
interesting in the point that
its reheating temperature $T_R$ is expected to be
sufficiently low to avoid the overproduction of gravitinos in
supergravity~\cite{Ellis}.  In particular, the gauge-mediated
supersymmetry(SUSY)-breaking model~\cite{Dine,Intriligator,Haba}
predicts the mass of gravitino $m_{3/2}$ in a range of $10^2$keV --
1GeV\cite{Gouvea}.  In this case the upper bounds of the reheating
temperature should be $T_R < 10^2$GeV -- $10^6$GeV for $m_{3/2} \simeq
10^{2} {\rm keV} - 1 {\rm
GeV}$~\cite{Moroi}.\footnote{
If $m_{3/2} \lesssim 1$keV, we have no constraint on the reheating
temperature. There have been found a few models~\cite{Strong,Izawa}
accommodating such a light gravitino, so far.}
This constraint is easily satisfied in a large class of `new
inflation' models.

However, the `new inflation' model has a serious drawback.
It requires an extreme fine tuning of the initial
condition~\cite{Linde-book} to have a sufficiently long inflation:
the universe has to have a large region over horizons at the beginning
where the inflation field $\varphi$ is smooth and its average value is
very close to a local maximum of the potential $V(\varphi)$. Since the
inflaton potential $V(\varphi)$ should be very flat to satisfy the
slow-roll condition, there is no dynamical reason for the
universe to choose such a specific initial value of the $\varphi$.

In this paper we point out that if there exists another inflation
(`pre-inflation') before the `new inflation', the required initial
condition for the `new inflaton' $\varphi$ is dynamically realized
owing to supergravity effects.\footnote{
In a large class of `new inflation' models the Hubble parameter is
much smaller than the gravitational scale. Thus, the `new inflation'
itself does not give a perfect explanation for why our universe lived for
such a long time. To solve this problem the presence of the
`pre-inflation' with a large Hubble parameter is
desirable~\cite{Linde-book}.}
To demonstrate our point, we give an explicit model for both of the
`new inflation' and the `pre-inflation'.  We also construct a
dynamical model which produces the desired potential for `new
inflation'.

\section{A `new inflation' model}

Our model consists of two inflaton potentials: one is for a `new
inflation' and the other for a `pre-inflation'. In this section we
consider the part of the `new inflation'. We adopt a `new
inflation' model proposed recently in Ref.~\cite{IzYa}, which is based
on an $R$ symmetry in supergravity.

The inflaton superfield $\phi(x, \theta)$ is assumed to have an $R$
charge $2/(n+1)$ so that the following tree-level superpotential is
allowed:
\begin{equation}
        W_{0}(\phi) = -\frac{g}{n+1}\phi^{n+1},
        \label{sup-pot}
\end{equation}
where $n$ is a positive integer and $g$ denotes a coupling constant of order
one. Here and hereafter, we set the gravitational scale $M\simeq
2.4\times 10^{18}$GeV equal to unity and regard it as a plausible
cutoff in supergravity. We further assume that the continuous $U(1)_R$
symmetry is dynamically broken down to a discrete $Z_{2nR}$ at a scale
$v$ generating an effective superpotential~\cite{IzYa,Kumekawa}:
\begin{equation}
        W_{eff}(\phi) = v^{2}\phi - \frac{g}{n+1}\phi^{n+1}.
        \label{sup-pot2}
\end{equation}
An explicit dynamics to induce this superpotential will be given in
section~4.

The $R$-invariant effective K\"ahler potential is given by
\begin{equation}
        K(\phi) = |\phi|^{2} + \frac{k}{4} |\phi|^{4} + \cdots,
        \label{kahler}
\end{equation}
where $k$ is a constant of order one.

The effective potential for a scalar component of the superfield
$\phi(x,\theta)$ in supergravity is given by~\cite{Nilles}
\begin{equation}
        V = e^{K(\phi)}\left\{ \left(\frac{\partial
        K}{\partial\phi\partial\phi^{*}}\right)^{-1}|D_{\phi}W|^{2}
        - 3 |W|^{2}\right\},
        \label{pot}
\end{equation}
with
\begin{equation}
        D_{\phi}W = \frac{\partial W}{\partial \phi}
        + \frac{\partial K}{\partial \phi}W,
        \label{DW}
\end{equation}
where $\phi(x)$ denotes a scalar component of $\phi(x,\theta)$. This
potential yields a vacuum
\begin{equation}
        \langle \phi \rangle \simeq
        \left(\frac{v^{2}}{g}\right)^{\frac{1}{n}},
\end{equation}
in which we have a negative energy as
\begin{equation}
        \langle V \rangle \simeq -3 e^{\langle K \rangle}
        |\langle W \rangle |^{2}
        \simeq -3 \left( \frac{n}{n+1}\right)^{2}|v|^{4}
        |\langle \phi \rangle|^{2}.
        \label{vacuum}
\end{equation}

It is a very interesting assumption in Ref.~\cite{IzYa} that the
negative vacuum energy (\ref{vacuum}) is canceled out by a
SUSY-breaking effect which gives a positive contribution
$\Lambda^{4}_{SUSY}$ to the vacuum energy. Namely, we impose
\begin{equation}
        -3\left(\frac{n}{n+1}\right)^{2}|v|^{4}
        \left|\frac{v^{2}}{g}\right|^{\frac{2}{n}}
        + \Lambda^{4}_{SUSY} = 0.
\end{equation}
In supergravity the gravitino acquires a mass
\begin{equation}
        m_{3/2} \simeq \frac{\Lambda^{2}_{SUSY}}{\sqrt{3}}
        = \left(\frac{n}{n+1}\right) |v|^{2}
        \left|\frac{v^{2}}{g}\right|^{\frac{1}{n}}.
        \label{gravitino-mass}
\end{equation}

The inflaton $\phi$ has a mass $m_{\phi}$ in the vacuum with
\begin{equation}
        m_{\phi} \simeq n |g|^{\frac{1}{n}}|v|^{2-\frac{2}{n}}.
        \label{inftaton-mass}
\end{equation}
The inflaton $\phi$ may decay into the ordinary particles through
$R$-invariant interactions with the ordinary light fields $\psi_{i}$ in
the K\"ahler potential. In general, we have the following
interactions:
\begin{equation}
        K(\phi,\psi_{i}) = \sum k_{i}|\phi|^{2}|\psi_{i}|^{2}.
        \label{interaction}
\end{equation}
With these interactions the reheating temperature is given by (see
Ref.~\cite{IzYa} for details)
\begin{equation}
        T_{R} \simeq 0.1 n^{\frac{3}{2}}|g|^{\frac{2}{n+1}}
        m_{3/2}^{\frac{3n-1}{2(n+1)}},
        \label{Rtemp}
\end{equation}
for $k_i \simeq 1$. Requiring $T_{R}\gtrsim 10$MeV, we get a constraint
on $m_{3/2}$:
\begin{equation}
        m_{3/2} \gtrsim n^{-\frac{n+1}{3n-1}}
        |g|^{\frac{4}{3n-1}}10^{\frac{39(n+1)}{3n-1}}.
        \label{gravitino-const}
\end{equation}
For $g\simeq 1$ and $n=4$, we have\footnote{
It has been shown in Ref.~\cite{IzYa} that $n=4$ is the most plausible
case. }
\begin{equation}
        m_{3/2} \gtrsim 10^{-18} (\simeq 3 {\rm GeV}).
\end{equation}
This already seems to exclude the gauge-mediated SUSY-breaking
model in the present scenario.
However, if a pair of Higgs doublets $H$ and $\bar{H}$ in the
SUSY standard model has a suitable $U(1)_{R}$ charge, we may have a
K\"ahler interaction with
\begin{equation}
        K(\phi, H, \bar{H}) = h\phi^{*}H\bar{H} + h.c.
\end{equation}
Then we obtain the reheating temperature
\begin{equation}
        T_{R} \simeq 0.1 m_{\phi}^{3/2} \simeq
        0.1 n^{\frac{3}{2}}
        \left(\frac{n+1}{n}\right)^{\frac{3(n-1)}{2(n+1)}}
        |g|^{\frac{3(n^2+1)}{2n(n+1)}}
        m_{3/2}^{\frac{3(n-1)}{2(n+1)}}.
\end{equation}
The requirement $T_{R}\gtrsim 10$MeV leads to
\begin{equation}
        m_{3/2} \gtrsim 3\times 10^{-23}  (\simeq 0.07 {\rm MeV}),
\end{equation}
for $g\simeq 1$ and $n=4$. We assume this case in this paper since it
accommodates the light gravitino ($m_{3/2} \lesssim 1$GeV) predicted
in the gauge-mediated SUSY-breaking model.\footnote{
As mentioned in the introduction, we do not have any cosmological
constraints if $m_{3/2} \lesssim 1$keV~\cite{Primack}.}

Let us now discuss the inflationary dynamics of our `new inflation'
model.
We identify the inflaton field $\varphi(x)/\sqrt{2} (\ge 0)$
with the real  part of the field $\phi(x)$. The potential
for the inflaton is given by
\begin{equation}
    V(\varphi) \simeq v^4 - \frac{k}{2}v^4\varphi^2
    -\frac{g}{2^{\frac{n}{2}-1}}v^2\varphi^n
    + \frac{g^2}{2^n}\varphi^{2n},
\end{equation}
for $\varphi < \langle\varphi \rangle = \sqrt{2}\langle \phi \rangle$.
Here, $g$ and $v$ are taken to be positive. Notice that the
$k$-independent contribution of $\varphi^2$ term in $e^K|D_{\phi}W|^2$
is exactly canceled by that in $-3|W|^2$ as stressed
in Ref.~\cite{Kumekawa,Copeland}.

It has been shown in Ref.~\cite{IzYa} that the slow-roll condition
for the inflaton is satisfied for $k < 1$ and
$\varphi \lesssim \varphi_f$ where
\begin{equation}
   \varphi_f \simeq \sqrt{2}\left(\frac{(1-k)v^2}{gn(n-1)}
   \right)^{\frac{1}{n-2}}.
\end{equation}
This provides the value of $\varphi$ at the end of inflation.
The Hubble parameter during
the inflation ($0 < \varphi \lesssim \varphi_f$) is given by
\begin{equation}
    H\simeq \frac{\sqrt{V}}{\sqrt{3}} \simeq
    \frac{v^2}{\sqrt{3}}.
\end{equation}

As shown in Ref.~\cite{IzYa}, the spectral index $n_s$ of the density
fluctuations is given by
\begin{equation}
        n_{s} \simeq 1 - 2k.
\end{equation}
By using the experimental constraint $n_{s} \gtrsim 0.6$~\cite{Adams,COBE},
we take $k \lesssim 0.2$. Since there is no symmetry reason to suppress
the $k$ term, we assume $k \gtrsim 0.02$. With $0.02 \lesssim k
\lesssim 0.2$, we obtain the $e$-fold number $N$ as
\begin{eqnarray}
    N & \simeq & \int^{\varphi_{N}}_{\tilde{\varphi}}
    d\varphi \frac{v^{4}}{-kv^{4}\varphi} +
    \int_{\varphi_{f}}^{\tilde{\varphi}}
    d\varphi \frac{v^{4}}{-n g2^{-\frac{n-2}{2}}v^{4}\varphi^{n-1}}
    \nonumber \\
    & = &  \frac{1}{k}
    \ln\left(\frac{\tilde{\varphi}}{\varphi_{N}}\right)
    + \frac{1-nk}{(n-2)k(1-k)},
    \label{N-efold}
\end{eqnarray}
where
\begin{eqnarray}
    \varphi_{N} & \simeq &  \tilde{\varphi}e^{-k\bar{N}},
    \\
    \tilde{\varphi} & = & \sqrt{2}
    \left(\frac{kv^{2}}{gn}\right)^{\frac{1}{n-2}},
    \\
    \bar{N} & = & N- \frac{1-nk}{(n-2)k(1-k)}.
\end{eqnarray}
We use the $e$-fold number $N$ of the present horizon~\cite{Linde-book}
\begin{eqnarray}
    N & \simeq & 66 + \ln (H) + \frac{1}{3}\ln(T_R)
    - \frac{2}{3}\ln(m_{\varphi}) - \frac{2}{3}\ln(\langle\varphi\rangle)
    \nonumber \\
    & \simeq & 50,
\end{eqnarray}
to get $\varphi_N \simeq (10^{-5}-0.1)v$ for $0.02 \lesssim k \lesssim
0.2$, $g \simeq 1$ and $n = 4$.

The amplitude of primordial density fluctuations $\delta \rho/\rho$
is given by
\begin{equation}
    \frac{\delta\rho}{\rho} \simeq \frac{1}{5\sqrt{3}\pi}
    \frac{V^{3/2}(\varphi_{N})}{|V'(\varphi_{N})|}
    = \frac{1}{5\sqrt{3}\pi} \frac{v^{2}}{k\varphi_{N}}.
\end{equation}
\hspace{0cm}From the COBE normalization~\cite{COBE}
\begin{equation}
    \frac{V^{3/2}(\varphi_{N})}{|V'(\varphi_{N})|}
    \simeq 5.3\times 10^{-4},
\end{equation}
We obtain
\begin{eqnarray}
    v & \simeq & 3.7\times 10^{-4}k^{3/2}g^{-1/2}e^{-k\bar{N}}
    \nonumber\\
    & = & 1.8 \times 10^{-9} - 6.8\times 10^{-7},
    \\
    m_{3/2} &\simeq & 2.7\times 10^{-9}k^{15/4}g^{-3/2}
    e^{-5k\bar{N}/2}
    \nonumber\\
    & = & 0.35 {\rm MeV} - 9.3\times 10^{2}{\rm GeV},
\end{eqnarray}
for $0.02 \lesssim k \lesssim 0.2$ and $g \simeq 1$. Here we have assumed
$n=4$.

We find that our `new inflation' model can accommodate the gravitino
in a large range of masses $m_{3/2} \simeq 0.3{\rm MeV} - 1{\rm
TeV}$ with the reheating temperature $T_{R} \simeq 40{\rm MeV} -
2\times 10^{4} {\rm GeV} $ and $n_{s} \simeq 0.6 - 0.96$.\footnote{
If one imposes $m_{3/2} \lesssim 1{\rm GeV}$
as suggested from the gauge-mediated SUSY-breaking
model~\cite{Dine,Intriligator,Haba}, one predicts $n_s \lesssim 0.75$,
which may be tested in future observations on the microwave background
radiation.}
However, we have just assumed the initial value of $\varphi \lesssim
\varphi_N \simeq (10^{-5}-0.1)v$, so far. In the next section we show that
another inflation (`pre-inflation') before the `new inflation'
naturally sets the initial value of the $\varphi \lesssim \varphi_N$
through supergravity effects.

\section{A `pre-inflation' model}

In this section we discuss another inflation before the `new
inflation' and show that the initial value required for the
$\varphi(x)$ is dynamically tuned during the `pre-inflation'. We
adopt a hybrid inflation model in Ref.~\cite{hybrid}
as an example of the `pre-inflation'.

The hybrid inflation model contains two kinds of superfields: one is
$S(x,\theta)$ and the others are $\Psi(x,\theta)$ and
$\bar{\Psi}(x,\theta)$. The model is also based on the $U(1)_R$
symmetry. The superpotential is given by~\cite{Copeland,hybrid}\footnote{
Symmetries of this model are discussed in Ref.~\cite{Copeland,hybrid}.}
\begin{equation}
    W = -\mu^{2} S + \lambda S \bar{\Psi}\Psi.
\end{equation}
The $R$-invariant K\"ahler potential is given by
\begin{equation}
    K(S,\Psi,\bar{\Psi}) = |S|^{2} + |\Psi|^{2} + |\bar{\Psi}|^{2}
    -\frac{k'}{4}|S|^{4} + \cdots ,
\end{equation}
where the ellipsis denotes higher-order terms, which we neglect in the
present analysis.

The potential in supergravity is given by
\begin{equation}
    V \simeq |\mu^{2} - \lambda\bar{\Psi}\Psi|^{2}
    + |\lambda \Psi S|^{2} + |\lambda\bar{\Psi}S|^{2}
        + k'\mu^{4}|S|^{2} + \cdots,
\end{equation}
where scalar components of the superfields are denoted by the same
symbols as the corresponding superfields. We readily see that if
the universe starts with sufficiently large value of $S$, the inflation
occurs for $0< k' <1$ and continues until $S\simeq S_{c}=
\mu/\sqrt{\lambda}$. The inflaton field $\sigma/\sqrt{2}$ is identified with
the real part of $S(x)$. The potential
is written as
\begin{equation}
    V\simeq |\mu^{2} - \lambda \bar{\Psi}\Psi |^{2}
    + \frac{|\lambda|^{2}}{2}\sigma^{2}(|\Psi|^{2}+|\bar{\Psi}|^{2})
    + \frac{k'}{2}\mu^{4}\sigma^{2} + \cdots.
\end{equation}
The Hubble parameter and $e$-folding factor $N'$ are given by
\begin{eqnarray}
    H & \simeq & \frac{\mu^{2}}{\sqrt{3}},
    \\
    N'& \simeq  & \frac{1}{k'}
    \ln\frac{\lambda\sigma_{i}^{2}}{2\mu^{2}}.\label{Ndash}
\end{eqnarray}

The point is that the `pre-inflation' implies a dynamical
tuning of the initial condition for the `new inflation'
in supergravity. During the `pre-inflation' the
inflaton field $\varphi(x)$ that causes the `new inflation' gets
an effective mass $\mu^2$ from the $e^{K}V$
term~\cite{Gaillard}. Since this effective mass ($\simeq \sqrt{3}H$)
is larger than the Hubble parameter during the `pre-inflation',
the $\varphi$ oscillates with its amplitude decreasing as $a^{-3/2}$
where $a$ denotes the scale factor.
Thus, at the end of the `pre-inflation', the
$\varphi$ takes a value
\begin{equation}
    \varphi \simeq \varphi_{i} e^{-\frac{3}{2}N'},
\end{equation}
where $\varphi_{i}$ is the value of $\varphi$ at the beginning of the
`pre-inflation'. We take $\varphi_{i}\sim 1$. Requiring $\varphi
\lesssim \varphi_{N}$, we get
\begin{equation}
    N' \gtrsim 10 - 20,
    \label{pre-N}
\end{equation}
for $n=4$, $g \simeq 1$ and $k = 0.02 - 0.2$.  Eq.(\ref{pre-N}) suggests
$\sigma_{i} \gtrsim (3.8 - 10)\mu/\sqrt{\lambda}$ for $k'\simeq 0.1$,
which is consistent with the assumption $\sigma_{i} \lesssim 1$
even for $\mu \simeq 10^{-2}$. \footnote{
As pointed out in Ref.~\cite{Dvali}, one-loop correction to the
potential is sizable if the coupling $\lambda$ is large ($\lambda
\gtrsim 4.3 k'^{1/3}\mu^{2/3}$). When the one-loop correction controls
the slow-roll dynamics, the condition $N' \gtrsim 10-20$ leads to
$\sigma_{i} \gtrsim (0.5-0.7) \lambda$, which is still consistent with
$\sigma_{i} \lesssim 1$ as far as $\lambda \lesssim 1$.}

However, the above condition is not sufficient since the minimum of
the potential for $\varphi$ deviates from zero through the effect of
$|D_{S}W|^2 + |D_{\phi}W|^2 -3|W|^2$ term.
The effective potential for $\varphi$ during the
`pre-inflation' is written as\footnote{
For simplicity, we neglect the $|\phi|^{2}|S|^{2}$ term in the K\"ahler
potential.}
\begin{equation}
    V(\varphi) \simeq \frac{1}{2} \mu^{4}\varphi^{2}+
    \frac{\sqrt{2}}{\sqrt{\lambda}}v^{2}\mu^{3}\varphi
    + \cdots.
\end{equation}
This potential has a minimum
\begin{equation}
    \varphi_{\rm min} \simeq -\frac{\sqrt{2}}{\sqrt{\lambda}}
    v\left(\frac{v}{\mu}\right).
\end{equation}
For a successful `new inflation' this minimum should be less than
$\varphi_{N}$. Therefore, we obtain the constrain on $\mu$ as $\mu
\gtrsim (10^{2}-10^{6})v$ for $\lambda \simeq 0.1$.

So far, we have seen that if $\mu \gtrsim (10^{2}-10^{6})v$ and
$\sigma_{i} \gtrsim (12 - 32) \mu$ for $\lambda \simeq 0.1$ and $k'
\simeq 0.1$, the inflaton field $\varphi$ for the `new inflation'
takes a value consistent with the initial condition for the `new
inflation' at the end of the `pre-inflation'.

We now discuss quantum effects during the `pre-inflation'. It is known
that in the de Sitter universe massless fields have quantum
fluctuations whose amplitude is given by $H/(2\pi)$. If the
fluctuations of the inflaton $\varphi$ was larger than $\sim
(10^{-5}-0.1)v$, we would yet have the initial value problem.
Fortunately, since the mass of the $\varphi$ during the
`pre-inflation' is not less than the Hubble parameter,
the quantum fluctuations for $\varphi$ are strongly
suppressed~\cite{Enquvist}. In fact,
the quantum fluctuations for $\varphi$ with wavelength corresponding to the
horizon at the beginning of the `new inflation' are given by
$ \delta \varphi \simeq
(H/2\pi)\exp[-(3/2)(N'-\ln(\mu/v))]$. Requiring $\delta\varphi
\lesssim (10^{-5}-0.1)v$ and using Eq.(\ref{Ndash}) we obtain a
constraint, $\mu \lesssim (4-9)\times 10^{-2}$, for $k'
\simeq 0.1, \lambda \simeq 0.1$ and $\sigma_i \lesssim 1$.  Therefore,
the dynamical tuning of the initial value of $\varphi$ discussed in
this section does work even if the Hubble parameter $H =
\mu^2/\sqrt{3}$ is larger than $v$.

\section{A dynamical model generating the potential for the `new inflation'}

We have two independent scales $v\simeq 10^{-9}-10^{-6}$ and $10^{-3} -
10^{-4} \lesssim \mu \lesssim 4\times 10^{-2}-9\times 10^{-2}$ as the
scales of the `new inflation' and the `pre-inflation',
respectively. We may identify the scale of the `pre-inflation' with
that of grand unification, $\mu \simeq 10^{-2}$, which may be related
to some fundamental physics in supergravity. On the other hand, it
is very natural to consider that the scale of the `new inflation'
arises from nonperturbative dynamics of a SUSY gauge theory, since it
is so small compared with the gravitational scale. In this section we
construct a model where the scale of the `new inflation' has the
dynamical origin.  We take $n=4$ for simplicity. Its generalization is
manifest.

Our model is based on a SUSY $SU(2)$ gauge theory with four doublet
hyperqaurks $Q^{\alpha}_{i}$ $(\alpha = 1,2, i=1,\cdots,4)$.
An effective superpotential~\cite{Seinberg} which describes the dynamics
of the $SU(2)$ gauge interaction may be given by
\begin{equation}
    W_{eff} = X({\rm Pf} V_{ij} - \Lambda^4)
    \label{dynamical-sup-potX}
\end{equation}
in terms of gauge-invariant degrees of freedom
\begin{equation}
    V_{ij} \sim Q_i Q_j,
\end{equation}
where the $X$ is an additional chiral superfield and Pf denotes
the Pfaffian. The global $SU(4)$ flavor symmetry is spontaneously
broken and massless Nambu-Goldstone fields are produced. To avoid
such massless fields, we explicitly break the global $SU(4)$ down to
$SP(4)$ by introducing an antisymmetric singlet superfield $Y^{ij}$
$(= -Y^{ji})$ which constitutes {\bf 5} of the global $SP(4)$.
Then we may have an $SP(4)$-invariant superpotential
\begin{equation}
    W = h Y^{ij} Q_i Q_j.
    \label{dynamical-sup-potY}
\end{equation}
Together with the superpotential Eq.(\ref{dynamical-sup-potX}),
we have a unique $SP(4)$-invariant vacuum
\begin{equation}
    \langle Q_1Q_2 \rangle=\langle Q_3Q_4 \rangle
    = \langle (QQ) \rangle = \Lambda^2; \quad
    (QQ) \equiv \frac{1}{2}(Q_1 Q_2 + Q_3 Q_4).
\end{equation}

We now introduce a singlet superfield $\phi(x,\theta)$.  We assume the
$U(1)_R$ charges of $Q^i$ and
$\phi$ to be $2/5$. Then we
obtain a $U(1)_R$-invariant superpotential\footnote{
In the limit of $g = 0$, the SUSY is
dynamically broken~\cite{VSB}.}
\begin{equation}
    W(Q,\phi) =  g'(QQ)^2 \phi
    - \frac{g}{5}\phi^5,
\end{equation}
where we have imposed a discrete symmetry $QQ \rightarrow -QQ$ and
$\phi \rightarrow \phi$.  The scale of the `new inflation' considered
in section~2 is given by
\begin{eqnarray}
    v^2 & = & g' \Lambda^4.
\end{eqnarray}
For $v\simeq 10^{-9}-10^{-6}$ and $g' \simeq 1$ we obtain $\Lambda \simeq
10^{-5} - 10^{-3} (\simeq 10^{13}-10^{15}{\rm GeV})$.

\section{Conclusion}

The `new inflation' model is the most attractive one among various
inflation models
in the point that it provides a reheating temperature low enough
to avoid the overproduction of gravitinos. However, the `new
inflation' requires an extreme fine tuning of the initial value of
the inflaton field $\varphi$. In this paper we have pointed out that for a
class of `new inflation' models\footnote{
It may be remarkable that the origin $\varphi = 0$ is a unique
$Z_{2nR}$-invariant point in the `new inflation' model in
Ref.~\cite{IzYa}. A model with a similar symmetry has been recently proposed
in Ref.~\cite{Dine2}.}
the required initial value of the
$\varphi$ is dynamically set if there is another inflation before the
`new inflation'.

Although we have used a hybrid inflation model for the
`pre-inflation' as an example, our observation is more general. Any
inflation that occurs before the `new inflation' can tune the
necessary initial value of the inflaton $\varphi$ ($\varphi \simeq 0$)
for the `new inflation' through the supergravity effects if the scale of
`pre-inflation' is sufficiently large.

\end{document}